\documentclass[a4paper,fleqn,usenatbib]{mnras}

\usepackage[T1]{fontenc}
\usepackage{ae,aecompl}

\usepackage{graphicx}	
\usepackage{amsmath}	
\usepackage{amssymb}	

\usepackage{graphicx,color}

\usepackage{blindtext}

\title[Self-similar solution of advection-dominated discs]{The self-similar structure of advection-dominated discs with outflow and radial viscosity}
\author[S. M. Ghoreyshi \& M. Shadmehri ]{
S. M. Ghoreyshi$^{1,2}$\thanks{E-mail: smghoreyshi64@gmail.com} and M. Shadmehri$^{1}$
\\
$^{1}$  Department of Physics, Faculty of Sciences, Golestan University, Gorgan 49138-15739, Iran\\
$^{2}$Research Institute for Astronomy and Astrophysics of Maragha (RIAAM), Maragha, P.O. Box: 55134-441, Iran\\
}

\date{Accepted 2020 February 24; Received 2020 February 12; in original form 2020 January 1}

\pubyear{2020}

\begin{document}
\label{firstpage}
\pagerange{\pageref{firstpage}--\pageref{lastpage}}
\maketitle

\begin{abstract}
Observational evidence and theoretical arguments postulate that outflows may play a significant role in the advection-dominated accretion discs (ADAFs). While the azimuthal viscosity is the main focus of most previous studies in this context, recent studies indicated that disc structure can also be affected by the radial viscosity. In this work, we incorporate these physical ingredients and the toroidal component of the magnetic field to explore their roles in the steady-state structure of ADAFs. We thereby present a set of similarity solutions where outflows contribute to the mass loss, angular momentum removal, and the energy extraction. Our solutions indicate that the radial viscosity causes the disc to rotate with a slower rate, whereas the radial gas velocity increases. For strong winds, the infall velocity may be of order the Keplerian speed if the radial viscosity is considered and the saturated conduction parameter is high enough. We show that the strength of magnetic field and of wind can affect the effectiveness of radial viscosity.
\end{abstract}



\section{Introduction}

In recent decades,  there is a growing interest to understand accretion processes in astrophysics  and different theoretical models for the accreting flows have been proposed. In the standard accretion disc model \citep{Shakura1973}, the viscous heating is balanced only by the radiative cooling and other cooling mechanisms such as the advective cooling are not considered. This model is valid when the mass accretion rate is sufficiently small and it fails for high and very low accretion rates \citep{Kato2008}. Depending on the accretion rate and the optical depth, the accretion discs with dominated advective cooling can be classified into two categories. If the accretion rate is significantly smaller than the Eddington rate, the accreting flow becomes optically thin and the disc is known as optically thin advection-dominated accretion flow (ADAF) \citep{Ichimaru1977,Narayan1994}. But when the accretion rate is much larger than the Eddington rate, the optical depth is very high and the radiated photons may be trapped inside the accretion flow. Such an optically thick advection-dominated disc is known as an optically thick ADAF or slim disc \citep{Abramowicz1988}. These models have been implemented to describe sources such as GRS 1915+105 \citep[e.g.,][]{Chen2004}, Seyfert 1 AGN \citep[e.g.,][]{Meyer2011}, and Sgr A* \citep[e.g.,][]{Yuan2002,Yuan2014}.

Recent observations of Sgr A* have shown that the mass accretion rate decreases inward \citep{Marrone2007,wang2013}. There are two models to explain this  trend of the accretion rate. In  the advection-dominated inflow-outflow solution (ADIOS), this reduction of the mass accretion rate is explained in terms of the  mass loss due to the outflow \citep[e.g.,][]{Blandford1999,Begelman2012}. But within the framework of the convection-dominated accretion flow (CDAF), the inward decrease of the mass accretion rate is resulted from locking gas to convective eddies moving in circular motion \citep[e.g.,][]{Narayan2000,Quataert2000,Abramowicz2002}. In the presence of the magnetic fields, however, \cite{Yuan2012a} showed that the hot accretion flow is not convectively unstable \citep[see also][]{Narayan2012}. Probably the CDAF scenario does not provide a plausible explanation for the accretion rate trend in Sgr A*.

The observations of other sources such as GRS 1915+105 have also shown that outflows may exist \citep[e.g.,][]{Neilsen2009,Miller2016}. The emergence of outflows in the advection-dominated discs has been confirmed by the numerical simulations \citep[e.g.,][]{Ohsuga2007,Takeuchi2009,Hashizume2015,Kitaki2018}. The outflows are driven by magnetic, thermal or radiative processes \citep[e.g.,][]{Chelouche2005,Cao2014,Ohsuga2011,Yuan2015,Hashizume2015} to extract mass, angular momentum, and energy from its host accretion disc \citep{Pudritz1985,Xue2005}. Disc structure, thereby, is significantly modified when contributions of the outflows are considered \citep[e.g.,][]{Blandford1982,Konigl1989}. For instance, \cite{Takeuchi2009} concluded that the surface density in inner regions of a slim disc strongly reduces due to the presence of outflows. As mentioned previously, the accretion rate in the presence of outflow also decreases inward \citep[e.g.,][]{Ohsuga2005,Yuan2012a,Bu2016}.

\cite{Blandford1999} constructed an ADAF model with outflow by assuming that the accretion rate is a power-law function of the radius, i.e., $\dot{M}\propto r^s$ where the exponent $s$ varies from 0 to 1. Note that the power law index, $s$, is a constant parameter and indicates the strength of the outflow. Not only the self-similar solutions, but only the global those have also shown that this relation may properly describe the influence of outflows on the disc structure \citep{Xie2008}. The numerical simulations of the hot accretion flows performed by \cite{Yuan2012a} also indicated that the accretion rate and the density may vary as power-law functions of the radial distance. Further theoretical studies have also shown that the power law index fits in a range of $0.5-1.0$ \citep[e.g.,][]{Stone1999,Ohsuga2005,Kawabata2009,Narayan2012,Yuan2012a,Yuan2012b,Bu2013,Yang2014,Bu2018}. In the case of Sgr A*, however, the radiatively inefficient accretion flow models imply that the power law index $s$ lies in a range between 0.3 to 0.4 \citep{Quataert1999,Yuan2003}.

Using self-similar solutions, the dynamical properties of the advection-dominated discs in the presence of outflows have been widely studied during recent years. Many authors studied properties of these discs in the presence of various physical ingredients like the magnetiec fields \citep[e.g.,][]{Bu2009,GhasemnezhadAbbassi2017}, magnetic diffusion \citep[e.g.,][]{Faghei2012}, self-gravity \citep[e.g.,][]{Shadmehri2009,Abbassi2013,Ghasemnezhad2016}, radial viscous force \citep[e.g.,][]{Beckert2000}, thermal conduction \citep[e.g.,][]{Shadmehri2008}, and convection \citep[e.g.,][]{Ghasemnezhad2017}. These studies led to interesting results. If the strong large-scale magnetic fields exist, for instance, the inflow of ADAFs in the presence outflows may be super-Keplerian and the temperature of the inflow decreases \citep{Bu2009}. Since the predicted temperature for ADAFs without either outflows or strong magnetic fields is higher than that obtained from fitting the observational data, such a lower temperature may be in good agreement with the observational values \citep{Yuan2004}.

We know that the ratio of the radial viscous force to the radial component of thermal pressure gradient for an accretion disc is proportional to $(H/r)^2$ in which $H$ is the disc half-thickness \citep{Frank2002}. Since the advection-dominated discs are geometrically thick, i.e., $H/r\sim1$, the radial viscosity may play a significant role in determining the properties of such discs (see also equation (\ref{eq:radial-force2})). The radial viscosity strongly affects the instabilities that may trigger in such discs \citep[see][]{Ghoreyshi2018}. \cite{Ghoreyshi2018} suggested that this type of viscosity explains the quasi-periodic oscillations in the black holes. To our knowledge, however, role of the radial viscosity in the structure of the advection-dominated discs with magnetic fields and the radial viscosity has not been studied so far. Although \cite{Beckert2000} presented similarity solutions for ADAFs including the radial viscosity, he did not consider the magnetic fields in his work \citep[see also][]{Narayan1995} and his main focus was on just the role of outflows. But \cite{Bu2009} showed that the magnetic fields and also the outflows can severely affect the disc structure and change previous results. In the present study, we shall investigate the properties of the advection-dominated discs including the outflows and the radial viscosity using similarity solutions. Our goal is to explore the effect of radial viscosity on the structure of the advection-dominated discs. In section 2, we formulate basic equations for a disc with outflows. Using self-similar method, we solve these equations and discuss our results in Section 3. We summarize our main findings and conclusions in the final section.


\section{Basic Equations}

We consider a cylindrical coordinate system $(r,\phi, z)$ that centered on the central object. The accretion disc is assumed to be axisymmetric ($\partial/\partial\phi=0$) and stationary ($\partial/\partial t=0$). The relativistic effects are neglected for simplicity and the Newtonian gravity is used. As we mentioned earlier, the advection-dominated discs are  geometrically thick, i.e., $H/r \leq 1$. We also assume that the dominant component of the magnetic field $\mathbf{B}$ is the toroidal component $B_{\phi}$ \citep[see][]{Hirose2004}. We also suppose that all flow variables depend only on the radial distance $r$. We use the formulation of \cite{Shadmehri2008} and \cite{Akizuki2006} and add the effect of the radial viscosity to their basic equations. The continuity equation is written as
\begin{equation}\label{eq:continuity}
\frac{d}{dr} (r \Sigma V_r)+\frac{1}{2\pi} \frac{d \dot{M}_W}{dr}=0,~~~~~~~~~~~~~~~~~~~~~~~~~~~~~~~~~~~~~~~~
\end{equation}
where $\Sigma$ is the surface density and is defined as $\Sigma=2\rho H$. Here, $\rho$ is the disc midplane density. The gas radial velocity is denoted by $V_r$ and its value is negative, i.e., $V_r<0$. The outflow mass-loss rate $\dot{M}_W$ is defined as
\begin{equation}\label{eq:mass-loss}
\dot{M}_W(r)=\int 4\pi r'\dot{m}_W(r') dr',~~~~~~~~~~~~~~~~~~~~~~~~~~~~~~~~~~~~~~~~
\end{equation}
where $\dot{m}_W$ is mass loss rate per unit area from each disc face.

By using the definition of the accretion rate $\dot{M}$ and its dependence on the radial distance \citep{Blandford1999}, we have
\begin{equation}\label{eq:accretion-rate}
\dot{M}=-2\pi r\Sigma V_r=\dot{M}_0(\frac{r}{r_0})^s,~~~~~~~~~~~~~~~~~~~~~~~~~~~~~~~~~~~~~~~~
\end{equation}
where $\dot{M}_0$ is the mass accretion rate at the outer boundary $r_0$. In the present paper, we suppose that typical values of the power law index $s$ vary from $s = 0$ to 0.3 (see the second paragraph of \S~\ref{sec S-F-S}). Using equations (\ref{eq:continuity}), (\ref{eq:mass-loss}), and (\ref{eq:accretion-rate}), we then arrive to this relation,
\begin{equation}
\dot{m}_W=\frac{s}{4\pi r_0^2}\dot{M}_0(\frac{r}{r_0})^{s-2}.~~~~~~~~~~~~~~~~~~~~~~~~~~~~~~~~~~~~~~~~
\end{equation}
One can see that for a given exponent $s$ the mass loss rate per unit area in the outer region is less than that in inner region.

The integrated radial momentum equation over $z$ with the radial viscous force becomes
\begin{eqnarray}\label{eq:radial-momentum}
V_r \frac{d V_r}{dr}=r(\Omega^2- {\Omega_K}^2)-\frac{1}{\Sigma}\frac{d}{dr}(\Sigma c_s^2)~~~~~~~~~~~~~~~~~~~~~~
 \nonumber\\-\frac{c_A^2}{r}-\frac{1}{2\Sigma}\frac{d}{dr}(\Sigma c_A^2)+F_\nu,
\end{eqnarray}
where $\Omega$ and $\Omega_K$ are the angular velocity and Keplerian angular speed on the equatorial plane, respectively. The Keplerian angular velocity is defined as $\sqrt{GM_\star/r^3}$ where $M_\star$ is the mass of the central object. The sound speed is ${c_s}=\sqrt{p/\rho}$ where $p$ is assumed to be equal to gas pressure, i.e., $p=p_{gas}$. We also define the Alfv\'{e}n as $c_A= B_\phi/ \sqrt{4\pi\rho}$.

The hydrostatic balance in the vertical direction leads to
\begin{equation}\label{eq:hydrostatic}
c_s=\frac{H\Omega_K}{\sqrt{1+\beta}},~~~~~~~~~~~~~~~~~~~~~~~~~~~~~~~~~~~~~~
\end{equation}
where $\beta$ is the ratio of magnetic field pressure to gas pressure and is defined as $(1/2)(c_A/c_s)^2$. This parameter is assumed to be constant through the disc and serves as an input parameter. Using different values of $\beta$, we can explore role of magnetic field in disc structure.

The radial viscous force $F_\nu$ is \citep{Papaloizou1986}
\begin{equation}\label{eq:radial-force}
F_\nu= \frac{1}{\Sigma}\frac{d}{dr} [\frac{4}{3}\frac{\nu_r\Sigma}{r}\frac{d}{dr}(rV_r)]-\frac{2V_r}{r\Sigma}\frac{d}{dr}(\nu_r\Sigma).
\end{equation}
Here, $\nu_r$ is the kinetic viscosity in the radial direction. If we consider molecular viscosity, $\nu_r$ may be equal to the viscosity associated with the azimuthal direction, i.e., $\nu$. If a turbulent viscosity is considered in the disc for which the turbulence is not necessarily isotropic, we cannot set $\nu_r=\nu$. In this paper, therefore, we assume that the ratio $\zeta=\nu_r/\nu$ is a given constant parameter. Note that the azimuthal viscosity is $\nu = \alpha c_s H$ where $\alpha$ is the turbulent coefficient with a value less than unity \citep{Shakura1973}.

In order to write the equation of motion in the azimuthal direction, we should obtain how much angular momentum is extracted by the
wind. The wind material is assumed to co-rotate with the disc. If the ejected material by the outflow is at radius $r$, the specific angular momentum carried away is $(\ell r)^2 \Omega$ \citep{Knigge1999}. Here $\Omega$ is the angular velocity of the disc at this radius and $\ell$ is the length of the rotational lever arm. Therefore, the vertically averaged azimuthal equation of motion is given by
\begin{equation}\label{eq:azimuthal-momentum}
 r\Sigma V_r \frac{d}{dr} (r^2 \Omega)=\frac{d}{dr}( r^3\nu\Sigma \frac{d \Omega}{dr})-\frac{\Omega(\ell r)^2}{2\pi} \frac{d\dot{M}_W}{dr}.~~~~~~~~~~~~~~~~
\end{equation}
Here, the angular momentum carried by the outflow is represented by last term of right hand side. Note that the type of outflow is parameterized in terms of a single parameter $\ell$. In a case with $\ell = 0$, the outflow is non-rotating \citep{Knigge1999} and does not extract any angular momentum, whereas a case with $\ell = 1$ corresponds to the specific angular-momentum-conserving disc winds. The later case is adequate for the radiation-driven outflows \citep{Proga1998}. In the cases with a wind parameter greater than unity, i.e. $\ell > 1$, the outflow can extract a lot of angular momentum from the disc. The centrifugally driven MHD winds \citep{Blandford1982} and the thermally driven winds \citep{Piran1977} are appropriate for this class.

We now consider the energy equation with the relevant cooling and heating mechanisms. We assume that the generated energy due to the viscous dissipation and the heat conduction are balanced by the advection cooling, radiative cooling and energy loss of the outflow. Thus, the energy equation becomes
\begin{eqnarray}\label{eq:energy}
\frac{1}{\gamma-1}\Sigma V_r\frac{d c_s^2}{dr}-2H V_r c_s^2\frac{d\rho}{dr}=Q_{vis}+Q_{vis_r}~~~~~~~~~~~~~~~
 \nonumber\\+Q_{cond}-Q_{rad}-Q_W,
\end{eqnarray}
where $Q_{vis}$, $Q_{vis_r}$, $Q_{cond}$, $Q_{rad}$, and $Q_W$ are the viscous heating in the azimuthal direction, viscous heating in the radial direction, thermal diffusion, radiative cooling, and the energy loss due to outflow, respectively. We employ the advection parameter $f =1 - \frac{Q_{rad}}{Q_{vis}}$ to measure the degree to which the accretion flow is advection-dominated \citep{Narayan1995}. Therefore, we can safely neglect radiative cooling for a case with $f \sim 1$. In this case, the disc is advection-dominated. For $f\ll1$, however, the disc is radiation-dominated. In general, $f$ is a function of $r$ and it depends on the details of heating and cooling processes. For simplicity, we assume that it is a constant given parameter. Thus, we can substitute $f Q_{vis}$ for $Q_{vis}-Q_{rad}$ in equation (\ref{eq:energy}). The viscous dissipation rates associated with the stresses in the azimuthal and radial directions, respectively, are written as \citep[see][]{Chen1993}
\begin{center}
    ${Q}_{vis}=\nu\Sigma(r\frac{d \Omega}{dr})^2,$
\end{center}
and
\begin{center}
    ${Q}_{vis_r}=2\nu_r\Sigma\Big\{(\frac{d V_r}{dr})^2+(\frac{V_r}{r})^2-\frac{1}{3}\big[\frac{1}{r}\frac{d}{dr}(rV_r)\big]^2\Big\}.$
\end{center}
The energy transported by conduction \citep{Cowie1977} and cooling due to outflow citep{Knigge1999} are
\begin{center}
    $Q_{cond}=-\frac{2H}{r}\frac{d}{dr}(rF_s),$
\end{center}
and
\begin{center}
    $Q_W=\frac{1}{2}\eta\dot{m}_W {V_K}^2.$
\end{center}
The saturated conduction flux is defined as $F_s =5\phi_s\rho {c_s}^3$, where $\phi_s$ is a constant of order unity \citep{Cowie1977}. Here, $V_K = r\Omega_K$ is the Keplerian speed. The dimensionless parameter $\eta$ quantifies the efficiency depending on the energy loss mechanisms. When this parameter is large, the extracted energy by outflow is larger \citep{Knigge1999}. Upon substituting  relations of ${Q}_{vis}$, ${Q}_{vis_r}$, $Q_{cond}$, and $Q_W$ into equation (\ref{eq:energy}), we obtain
\begin{eqnarray}\label{eq:main-energy}
\frac{1}{\gamma-1}\Sigma V_r\frac{d c_s^2}{dr}-2H V_r c_s^2\frac{d\rho}{dr}=f\nu\Sigma(r\frac{d \Omega}{dr})^2~~~~~~~~
\nonumber\\~~~ +2\nu\Sigma\Big\{(\frac{d V_r}{dr})^2+(\frac{V_r}{r})^2-\frac{1}{3}\big[\frac{1}{r}\frac{d}{dr}(r V_r)\big]^2\Big\}
  \nonumber\\~~~~~~~~~~~~~~~~~~~~~-\frac{2H}{r}\frac{d}{dr}(5r\phi_s\rho {c_s}^3)-\frac{1}{2}\eta\dot{m}_W {V_K}^2.
\end{eqnarray}

Finally, when the toroidal component of the magnetic field is dominant the induction equation can be written as \citep{Kato2008}
\begin{equation}\label{eq:induction}
\frac{d}{dr} (V_r B_\phi)=\dot{B}_\phi.~~~~~~~~~~~~~~~~~~~~~~~~~~~~~~~~~~~~~~~~
\end{equation}
where $\dot{B}_\phi$ denotes the field escaping/creating rate which may result from the magnetic diffusion or dynamo effect \citep{Machida2006,Oda2007}. By solving the basic equations (\ref{eq:continuity}), (\ref{eq:radial-momentum}), (\ref{eq:hydrostatic}), (\ref{eq:azimuthal-momentum}), (\ref{eq:main-energy}), and (\ref{eq:induction}), we investigate the dynamical properties of advection-dominated discs.


\section{Self-similar solutions}\label{sec S-F-S}

Although one can not discuss the global behaviour of an accretion flow by using self-similar solutions, the similarity solutions describe asymptotic behaviour of the accretion flow in the regions far from the inner and the outer disc boundaries. We assume that the disc quantities are power law functions of the radial distance. The exponents, thereby, are obtained self-consistently using basic equations. We suggest the following self-similar solutions:
\begin{equation}\label{eq:s}
\Sigma(r)=\omega_0\Sigma_0(\frac{r}{r_0})^{s-\frac{1}{2}},~~~~~~~~~~~~~~~~~~~~~~~~~~~~~~~~~~~~~~~~
\end{equation}

\begin{equation}\label{eq:o}
\Omega(r)=\omega_1\sqrt{\frac{GM_*}{r_0^3}}(\frac{r}{r_0})^{-3/2},~~~~~~~~~~~~~~~~~~~~~~~~~~~~~~~~~~~~~~~~
\end{equation}

\begin{equation}\label{eq:v}
V_r(r)=-\omega_2\sqrt{\frac{GM_*}{r_0}}(\frac{r}{r_0})^{-1/2},~~~~~~~~~~~~~~~~~~~~~~~~~~~~~~~~~~~~~~~~
\end{equation}

\begin{equation}\label{eq:p}
P(r)=\omega_3 \Sigma_0\frac{GM_*}{r_0}(\frac{r}{r_0})^{s-\frac{3}{2}},~~~~~~~~~~~~~~~~~~~~~~~~~~~~~~~~~~~~~~~~
\end{equation}

\begin{equation}\label{eq:cs}
c_s^2(r)=\frac{\omega_3}{\omega_0}\frac{GM_*}{r_0}(\frac{r}{r_0})^{-1},~~~~~~~~~~~~~~~~~~~~~~~~~~~~~~~~~~~~~~~~
\end{equation}

\begin{equation}\label{eq:ca}
c_A^2(r)=2\beta\frac{\omega_3}{\omega_0}\frac{GM_*}{r_0}(\frac{r}{r_0})^{-1},~~~~~~~~~~~~~~~~~~~~~~~~~~~~~~~~~~~~~~~~
\end{equation}

\begin{equation}\label{eq:h}
H(r)=\omega r_0(\frac{r}{r_0}),~~~~~~~~~~~~~~~~~~~~~~~~~~~~~~~~~~~~~~~~
\end{equation}

\begin{equation}\label{eq:b}
{\dot{B}_\phi}(r)=\dot{B}_{\phi0}(\frac{r}{r_0})^{(s-\frac{11}{2})/2},~~~~~~~~~~~~~~~~~~~~~~~~~~~~~~~~~~~~~~~~
\end{equation}
where $P$ is the height-integrated pressure. Furthermore, the reference quantities $\Sigma_0$, $r_0$ and $\dot{B}_{\phi0}$ are introduced to present the equations into dimensionless forms. Upon substituting the self-similar solutions (\ref{eq:s})$-$(\ref{eq:b}) into the basic equations (\ref{eq:continuity}), (\ref{eq:radial-momentum}), (\ref{eq:hydrostatic}), (\ref{eq:azimuthal-momentum}), (\ref{eq:main-energy}), and (\ref{eq:induction}), we have
\begin{equation}
\omega_0\omega_2=\dot{m},~~~~~~~~~~~~~~~~~~~~~~~~~~~~~~~~~~~~~~~~
\end{equation}

\begin{eqnarray}
-\frac{1}{2}{\omega_2}^2={\omega_1}^2-1-(s-\frac{3}{2})(1+\beta)\frac{\omega_3}{\omega_0}~~~~~~~~~~~~~~~~~~~~~
\nonumber\\-2\beta\frac{\omega_3}{\omega_0}+\frac{4}{3}\zeta\alpha(s+\frac{3}{4})\sqrt{1+\beta}\frac{\omega_3}{\omega_0}\omega_2,
\end{eqnarray}

\begin{equation}
\omega_0\omega^2-\omega_3(1+\beta)=0,~~~~~~~~~~~~~~~~~~~~~~~~~~~~~~~~~~~~~~~~
\end{equation}

\begin{equation}
\omega_0\omega_2=3\alpha(s+\frac{1}{2})\sqrt{1+\beta}\omega_3+2s\ell^2\dot{m},~~~~~~~~~~~~~~~~~~~~~~~~~~~~~~~
\end{equation}

\begin{eqnarray}
(\frac{1}{\gamma-1}+s-\frac{3}{2})\omega_2\omega_3=\frac{9}{4}f\alpha\sqrt{1+\beta}{\omega_1}^2\omega_3+\frac{7}{3}\zeta\alpha~~~~~~~
\nonumber\\\sqrt{1+\beta}{\omega_2}^2\omega_3-5\phi_s(s-2)\omega_0(\frac{\omega_3}{\omega_0})^{3/2}-\frac{1}{4}\eta s \dot{m},
\end{eqnarray}

\begin{equation}
\dot{B}_{\phi0}=G M_\star \sqrt{\frac{\pi\beta\Sigma_0}{{r_0}^5}}(\frac{7}{2}-s)\omega_2\sqrt{\frac{\omega_3}{\omega}}.~~~~~~~~~~~~~~~~~~~~~~~~~~~~~~~
\end{equation}
Here, $\dot{m}$ defined as $\dot{m}=\dot{M}_0/(2\pi r_0 \Sigma_0 \sqrt{GM_*/r_0})$ is the non-dimensional mass accretion rate. After mathematical manipulations, a fourth-order algebraic equation is obtained for $\omega$:
\begin{eqnarray}\label{eq:mainequation}
\frac{9\alpha^3(s+\frac{1}{2})}{(1-2s\ell^2)(1+\beta)}\Big[\frac{-9f(s+\frac{1}{2})}{8(1-2s\ell^2)}+\frac{7\zeta(s+\frac{1}{2})}{3(1-2s\ell^2)}~~~~~~~~~
\nonumber\\-f\zeta(s+\frac{3}{4})\Big]\omega^4+3\alpha\Big[\frac{3}{4}f(s-\frac{3}{2})+\frac{3}{2} f\frac{\beta}{1+\beta}~~~
\nonumber\\~~~~~~~~-\frac{s+\frac{1}{2}}{(1-2s\ell^2)(1+\beta)}\big(\frac{1}{\gamma-1}+s-\frac{3}{2}\big)\Big]\omega^2-5\phi_s
\nonumber\\~~~~~~~~~~~\frac{s-2}{1+\beta}\omega-\frac{3}{4}\alpha\eta s\frac{s+\frac{1}{2}}{1-2s\ell^2}+\frac{9}{4}f\alpha=0.
\end{eqnarray}
We can obtain other flow quantities as a function of $\omega$. Thus,
\begin{equation}\label{eq:sigma}
\omega_0=\frac{(1-2s\ell^2)\sqrt{1+\beta}}{3\alpha(s+\frac{1}{2})}\dot{m}\omega^{-2},~~~~~~~~~~~~~~~~~~~~~~~~~~~~~~~~~~~~~~~~
\end{equation}

\begin{eqnarray}
\omega_1=\Big\{\big[-\frac{9\alpha^2(s+\frac{1}{2})^2}{2(1+\beta)(1-2s\ell^2)^2}-\frac{4\zeta\alpha^2(s+\frac{1}{2})(s+\frac{3}{4})}{(1+\beta)(1-2s\ell^2)}\big]\omega^4
\nonumber\\~~~~~~~~~~~~~~~+\big[s-\frac{3}{2}+\frac{2\beta}{1+\beta}\big]\omega^2+1\Big\}^{1/2},
\end{eqnarray}

\begin{equation}\label{eq:radialvelosity}
\omega_2=\frac{3\alpha(s+\frac{1}{2})}{(1-2s\ell^2)\sqrt{1+\beta}}\omega^2,~~~~~~~~~~~~~~~~~~~~~~~~~~~~~~~~~~~~~~~~
\end{equation}

\begin{equation}\label{eq:pressure}
\omega_3=\frac{1-2s\ell^2}{3\alpha(s+\frac{1}{2})\sqrt{1+\beta}}\dot{m},~~~~~~~~~~~~~~~~~~~~~~~~~~~~~~~~~~~~~~~~
\end{equation}

\begin{equation}\label{eq:magneticf}
\dot{B}_{\phi0}=G M_\star \sqrt{\frac{\pi\beta\Sigma_0}{{r_0}^5}}(\frac{7}{2}-s)\big[\frac{3\alpha\dot{m}(s+\frac{1}{2})}{(1+\beta)^{3/2}(1-2s\ell^2)}\big]^{1/2}\omega^{3/2}.~~~~~~~~~~~~~~~~~~~~~~~~~~~~~~~
\end{equation}
In the absence of radial viscosity and for the nonmagnetic flows, the above solutions are reduces to those obtained by \cite{Shadmehri2008}. Properties of advection-dominated discs in the presence of radial viscosity are described by solving the equation (\ref{eq:mainequation}) numerically. Note that we adopt only real roots which correspond to ${\omega_1}^2 >0$. Equations (\ref{eq:sigma}), (\ref{eq:pressure}), and (\ref{eq:magneticf}) show that the surface density, the pressure $P$, and $\dot{B}_{\phi0}$ are directly proportional to the mass accretion rate $\dot{M}_0$. Other disc quantities, however, are not directly proportional to the accretion rate. Using these solutions, we find that the surface density decreases with increasing the viscosity parameter. However, this trend is different for the radial velocity (see equation (\ref{eq:radialvelosity})). These results were obtained by \cite{Shadmehri2008}.

\begin{figure}
\includegraphics[scale=0.4]{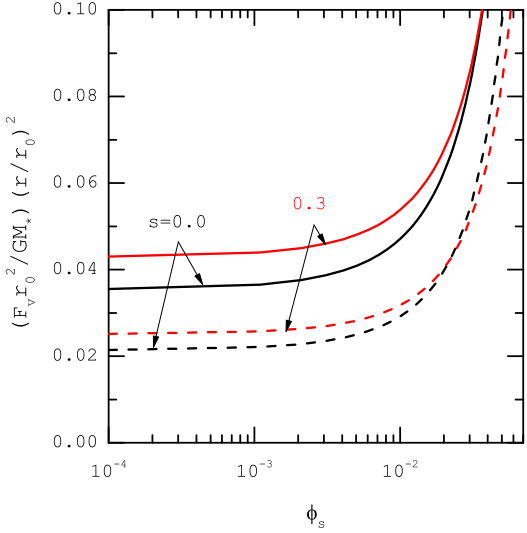}
\caption{The radial viscous force versus the saturation constant $\phi_s$ for different values of $s$, as labeled. The solid and dashed curves are the discs with $\beta=0.5$ and 0.125, respectively. The other model parameters are $\zeta=1.0$, $\alpha =0.2$, and $\ell=1.0$.}\label{fig:f1}
\end{figure}

For the numerical study of our model, we set $\dot{m}=0.1$, $\gamma=1.5$, $\zeta=1.0$, $\alpha=0.2$, $\beta=0.125$, and $f =\eta=\ell=1$, unless otherwise is stated. In each Figure, however, we adopt different values of $s$. As we mentioned earlier, \cite{Beckert2000} investigated the disc properties as a function of the viscosity parameter in the presence of radial viscous force. In the present paper, however, we study the disc properties, as \cite{Shadmehri2008} studied, versus the thermal conduction coefficient $\phi_s$. The acceptable range of parameter $\phi_s$ varies from $10^{-4}$ to 0.07. Large values of $\phi_s$ violate the restriction $H/r<1$, and thereby we don't consider higher value of the thermal conduction coefficient. On the other hand, the momentum conservation implies that the term $1-2s\ell^2$ must be greater than zero. Although this inequality is fulfilled for the discs with the non-rotating winds, i.e., $\ell=0$, this condition for the discs with the rotating winds leads to $s<1/2$. Because of the similarity approximation, our self-similar solutions for $s=0.4$ diverge and don't show the expected physical behavior. On the other hand, the value of $s$ can be sensitive to the value of $\alpha$ and increase as the $\alpha$ parameter decreases \citep{Yang2014}. \cite{Yang2014} found that for $\alpha=0.1$, for example, the power law index $s$ is 0.37. The observational studies have also shown that the power law index cannot be larger than 0.4 \citep{Yuan2003}. Therefore, we suppose that the value of $s$ may vary in the range of 0.0 (no outflow case) to 0.3 (a case with moderate outflow).

Using our self-similar solutions, the equation (\ref{eq:radial-force}) yields
\begin{equation}\label{eq:radial-force2}
F_\nu= 4\zeta\alpha^2\frac{GM_{\star}}{{r_0}^2}(\frac{r}{r_0})^{-2}\frac{(s+\frac{1}{2})(s+\frac{3}{4})}{(1+\beta)(1-2s\ell^2)} \omega^4.
\end{equation}
In Fig. \ref{fig:f1}, the radial viscous force versus the thermal conduction coefficient $\phi_s$ is illustrated for different values of $s$. In this figure, we suppose that $\beta$ is equal to 0.5, i.e., $c_A=c_s$, (solid curves) and 0.125, i.e., $c_A=c_s/2$, (dashed curves). We find that the radial force strongly depends on the wind strength and the magnetic field strength. One can see that the radial viscous force depends not only on the parameter $s$ and $\beta$, but also on the parameter $\phi_s$. In the range of $\phi_s=0.01-0.07$, the value of radial viscous force strongly increases. For lower values of $\phi_s$, however, the value of this force remains almost unchanged.

\begin{figure*}
\includegraphics[scale=1.0]{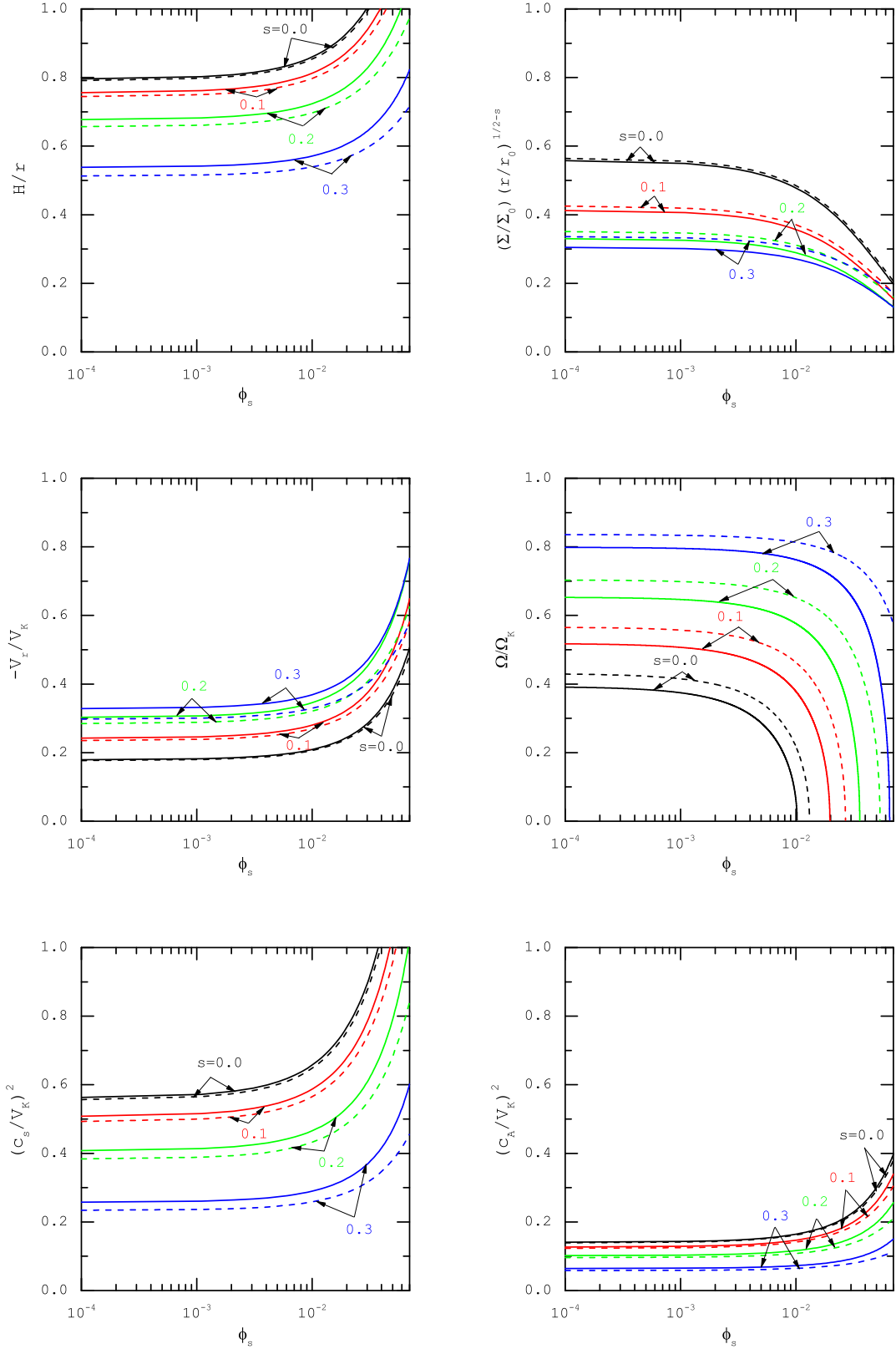}
\caption{Profiles of the physical variables of the accretion disc versus the saturation constant $\phi_s$, for $\zeta=1.0$, $\alpha =0.2$, $\beta=0.125$, and $\ell=1.0$. The solid and dashed curves correspond to the cases with and without the radial viscosity, respectively.}\label{fig:f2}
\end{figure*}

In Fig. \ref{fig:f2}, the profiles of physical variables are shown for $\beta=0.125$, $\zeta=1.0$, $\alpha=0.2$, and $\ell=1.0$. The solid and dash curves represent solutions with and without the radial viscosity, respectively. Each curve is labeled with the chosen value of exponent $s$. Top left panel shows the dimensionless ratio $H/r$. Although the disc thickness for stronger winds is smaller, the radial viscosity causes the disc to be thicker. For higher value of $\phi_s$, one can see that the disc thickness strongly increases. In top right panel, the dimensionless surface density versus $\phi_s$ is shown. As expected, the surface density is higher as the wind becomes weaker. This finding is in a good agreement with previous studies \citep[e.g.,][]{Shadmehri2008,Abbassi2013}. If the radial viscosity is ignored, the disc surface density becomes more. The surface density reduces as the conduction coefficient $\phi_s$ increases. Note that in the small-limit $\phi_s$, the reduction of surface density due to viscosity in the radial direction is less.

The radial velocity is illustrated in the middle left panel of Fig. \ref{fig:f2}. The inward motion of the disc material becomes faster when the radial viscosity is considered. We can ignore the influence of the radial viscosity in the small$-\phi_s$ limit when the wind is weak. If we consider the opposite limit, i.e., high values of $\phi_s$ and $s$, the viscosity in the radial direction plays an important role in the infall process. The angular velocity in a disc with $\beta=0.125$ is also shown in the middle right panel. We see that the radial viscosity causes the disc to rotate with a slower rate. We define the specific $\phi_s$ at which the accretion disc tends to a non-rotating limit no matter the radial viscosity is considered or not. In the non-magnetized case, a similar trend has already been obtained by \cite{Shadmehri2008}. For higher values of the specific $\phi_s$, ${\omega_1}^2$ is negative and the solutions are not physical. When the disc has, in addition to the azimuthal viscosity, a viscosity in the radial direction, it reaches a non-rotating limit at a lower value of $\phi_s$. But the wind causes the non-rotating limit to occur at higher values of $\phi_s$. In bottom panels, the speed sound (left panel) and the Alfv\'{e}n velocity (right panel) are displayed. Although the wind reduces the sound speed and the Alfv\'{e}n velocity, these speeds enhance due to the radial viscosity. Since ${c_A}^2$ (or ${c_s}^2$) is proportional to $\omega^2$ (see equations (\ref{eq:ca}), (\ref{eq:cs}), (\ref{eq:sigma}), and (\ref{eq:pressure})), the rise in the Alfv\'{e}n velocity (or the sound speed) is due to an increase in the ratio $H/r$.

\begin{figure*}
\includegraphics[scale=1.0]{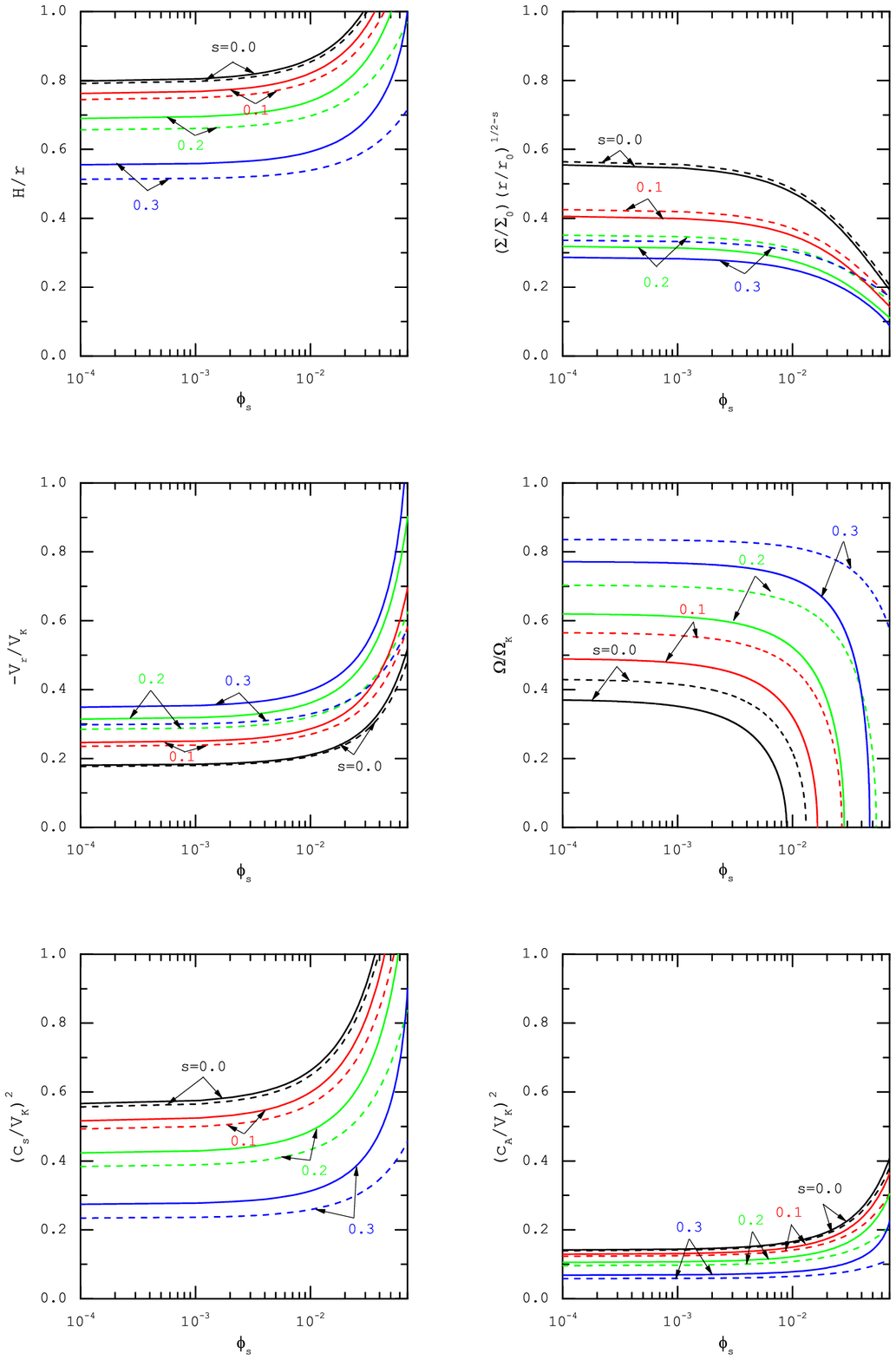}
\caption{Similar to Fig. \ref{fig:f2}, but for $\zeta=3/2$. The solid and dashed curves correspond to the cases with and without the radial viscosity, respectively.}\label{fig:f3}
\end{figure*}

In our study, the parameter $\zeta$ which quantifies the relative importance of the radial viscosity and the azimuthal viscosity plays a vital role. In Fig. \ref{fig:f3}, we study the effect of this parameter on the disc quantities. Although the $r\phi$-component of the stress tensor is usually considered to be the dominant component, the other components can paly an important role in transporting angular momentum \citep{Bai2013,Moeen2017}. Hence, the bulk viscosity would become comparable to the shear viscosity \citep[see also][]{Papaloizou1977}. Here, the radial viscosity is assumed to be 1.5 times the azimuthal viscosity, i.e., $\zeta=3/2$. Note that the previous works considered even higher values of $\zeta$ \citep[e.g.,][]{Papaloizou1986}. Other input parameters are similar to Fig. \ref{fig:f2}. As expected, a rise in $\zeta$ causes the role of radial viscosity to be more impressive. At high $\phi_s$-limit and for $s=0.3$, for example, the infall velocity in the presence of radial viscosity which tends to be of the order of the Keplerian speed is nearly two times that for a case without radial viscosity. Under these circumstances and for $zeta=1.0$, however, the ratio of these two infall velocities is about 1.3. The sound speed significantly increases due to the radial viscosity. By comparing Fig. \ref{fig:f3} and Fig. \ref{fig:f2}, we also find $H/r \simeq 1$ and the non-rotating limit for the disc are achieved at smaller $\phi_s$ if $\zeta$ increases.

\begin{figure*}
\includegraphics[scale=1.0]{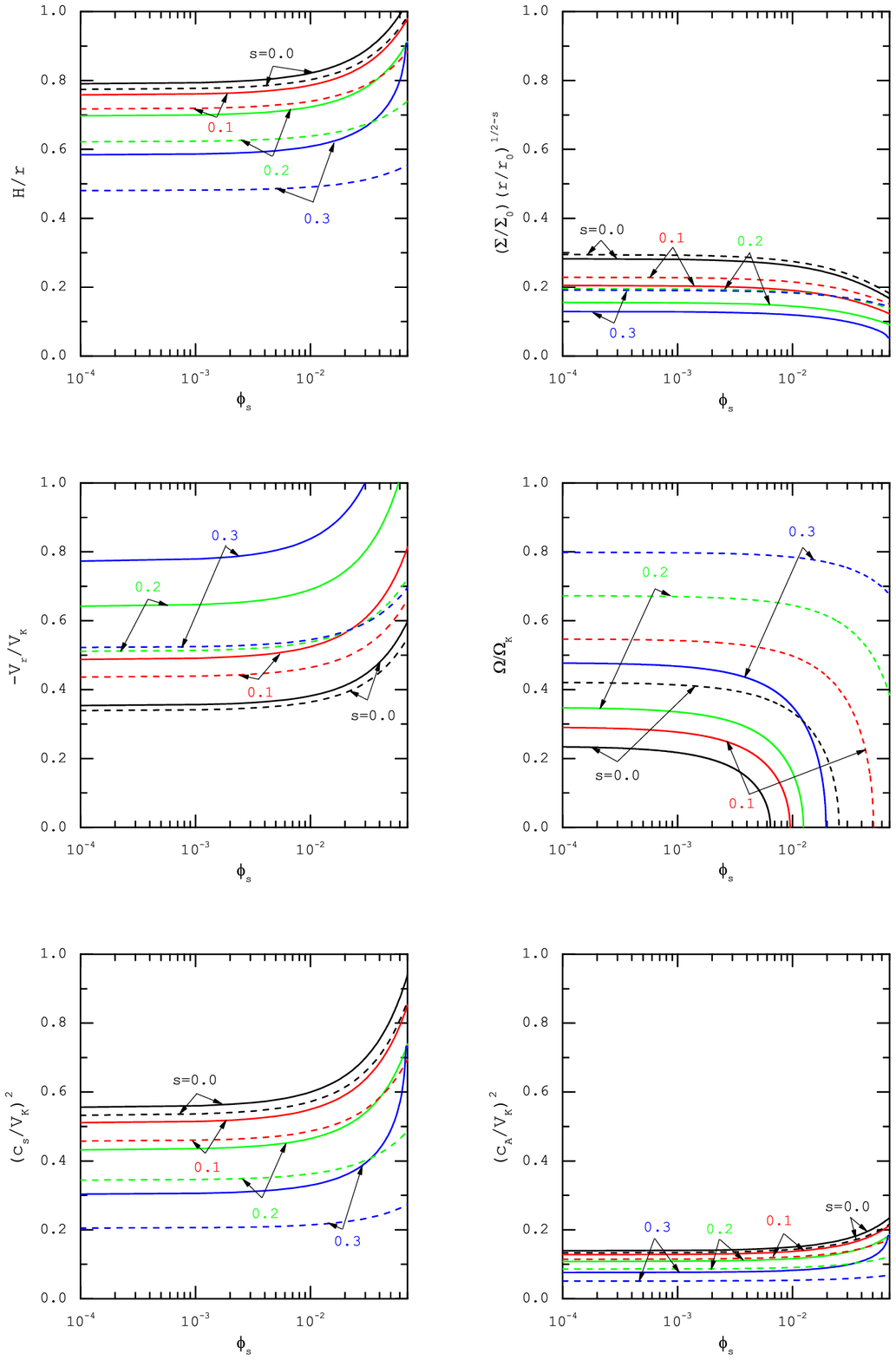}
\caption{Similar to Fig. \ref{fig:f2}, but for $\alpha=0.4$. The solid and dashed curves correspond to the cases with and without the radial viscosity, respectively.}\label{fig:f4}
\end{figure*}

The role of $\alpha$ parameter in the disc structure is displayed in Fig. \ref{fig:f4}. It is assumed that $\alpha =0.4$ and other parameters are similar to Fig. \ref{fig:f2}. Note that the change in $\alpha$ modifies the disc structure whether the radial viscosity is considered or not. We find that the disc thickness reduces in a case with higher value of $\alpha$ when the radial viscosity is absent. In the presence of radial viscosity, however, the modification of disc thickness depends on the wind strength $s$. By increasing the viscosity parameter, the reduction in surface density and rotational velocity is more obvious. As seen in the middle right panel of Fig. \ref{fig:f4}, the rotational velocity in the absence of radial viscosity is about two times that of a disc with the radial viscosity. Although an increase in the viscosity parameter causes the disc materials to rotate slower, their infall motion has a faster rate due to this rise. We find that not only the infall velocity, but also the sound speed of an ADAF with a moderate wind enhances.

\begin{figure*}
\includegraphics[scale=1.0]{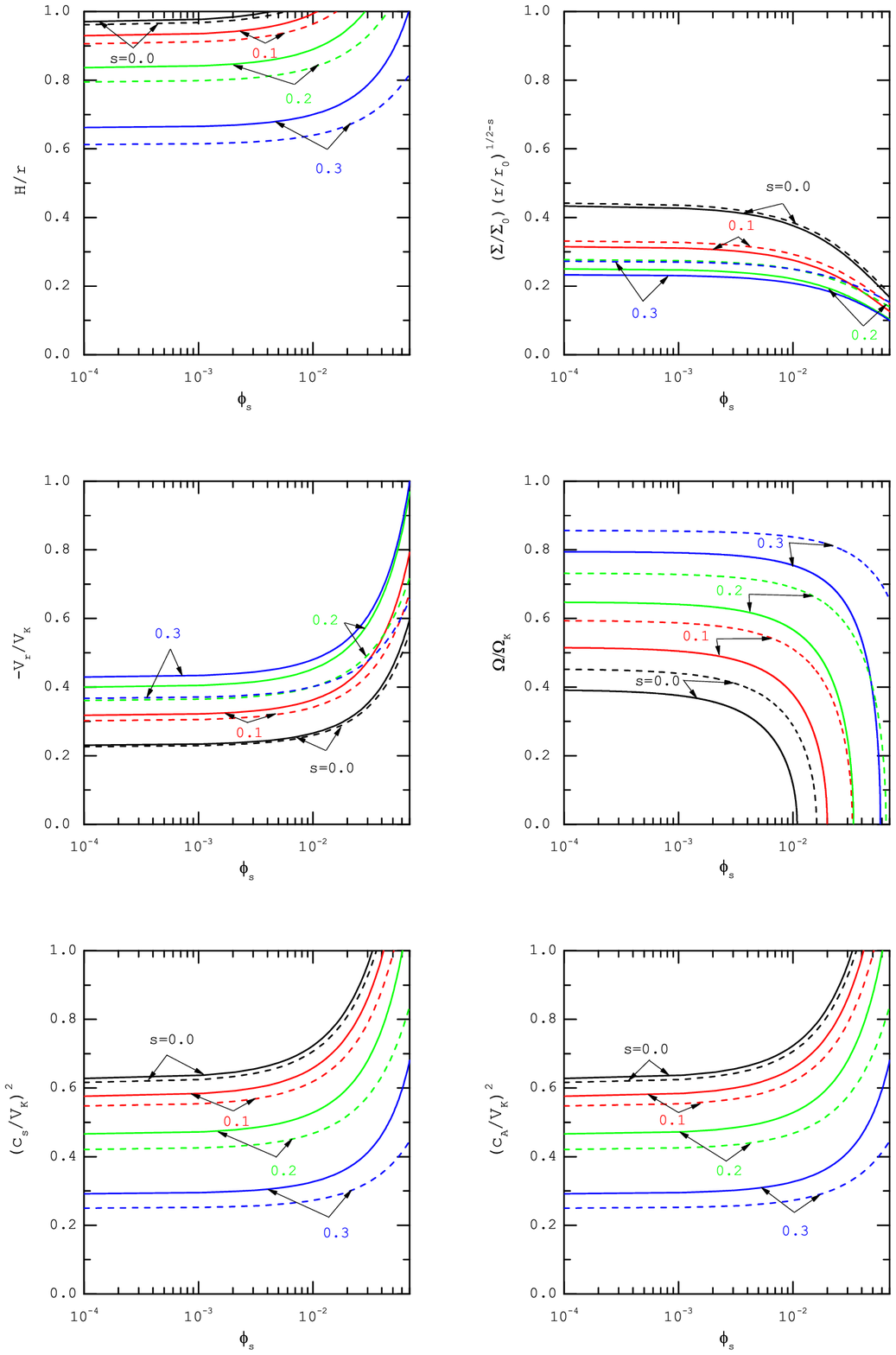}
\caption{Similar to Fig. \ref{fig:f2}, but for $\beta=0.5$. The solid and dashed curves correspond to the cases with and without the radial viscosity, respectively.}\label{fig:f5}
\end{figure*}

In Fig. \ref{fig:f5}, we show the role of the magnetic fields in the disc structure. Here, the Alfv\'{e}n velocity is assumed to be equal to the sound speed, i.e., $\beta=0.5$. The rest of the model parameters are similar to Fig. \ref{fig:f2}. By comparing Fig. \ref{fig:f5} with Fig. \ref{fig:f2}, one can find that a stronger magnetic field causes the disc to be thicker. This results is in good agreement with the findings of \cite{Ghasemnezhad2017} and \cite{GhasemnezhadAbbassi2017}. Note that their results show that the presence of vertical component of the magnetic field, i.e., $B_z$, causes the disc thickness to decrease. We find that the difference of disc thickness between the cases with and without the radial viscosity becomes more for stronger magnetic fields. This difference is related to the dependence of radial viscous force on $\beta$ (see Fig. \ref{fig:f1}). The disc material moves faster in the radial direction if the strength of magnetic field increases. The previous works have also demonstrated that an increase in not only the toroidal component, but also other those of the magnetic field yields a faster inflow \citep{Bu2009,Mosallanezhad2013}. One can see that this increase in the infall velocity depends on value of $s$. We find that, however, an increase in the value of $\beta$ causes the surface density to reduce which this reduction is also dependent upon the wind strength.

Although the angular velocity in the presence of radial viscosity is nearly independent of $\beta$, the rotational velocity of a disc without the radial viscosity enhances as the magnetic field becomes stronger. The value of $\phi_s$ for which the disc reaches to a non-rotating limit changes when only the azimuthal viscosity is considered. Note that the changes in the specific $\phi_s$ due to the magnetic field depend on the input parameter $s$. We also find that the sound speed and, in particular, the Alfv\'{e}n velocity strongly depends on the input parameters of $\beta$ and $s$. Although the toroidal magnetic field alone increases the disc temperature \citep[see also][]{Ghasemnezhad2016}, the presence of toroidal and poloidal components together implies that the temperature of disc reduces \citep[e.g.,][]{Bu2009,Mosallanezhad2013}.

\begin{figure*}
\includegraphics[scale=1.0]{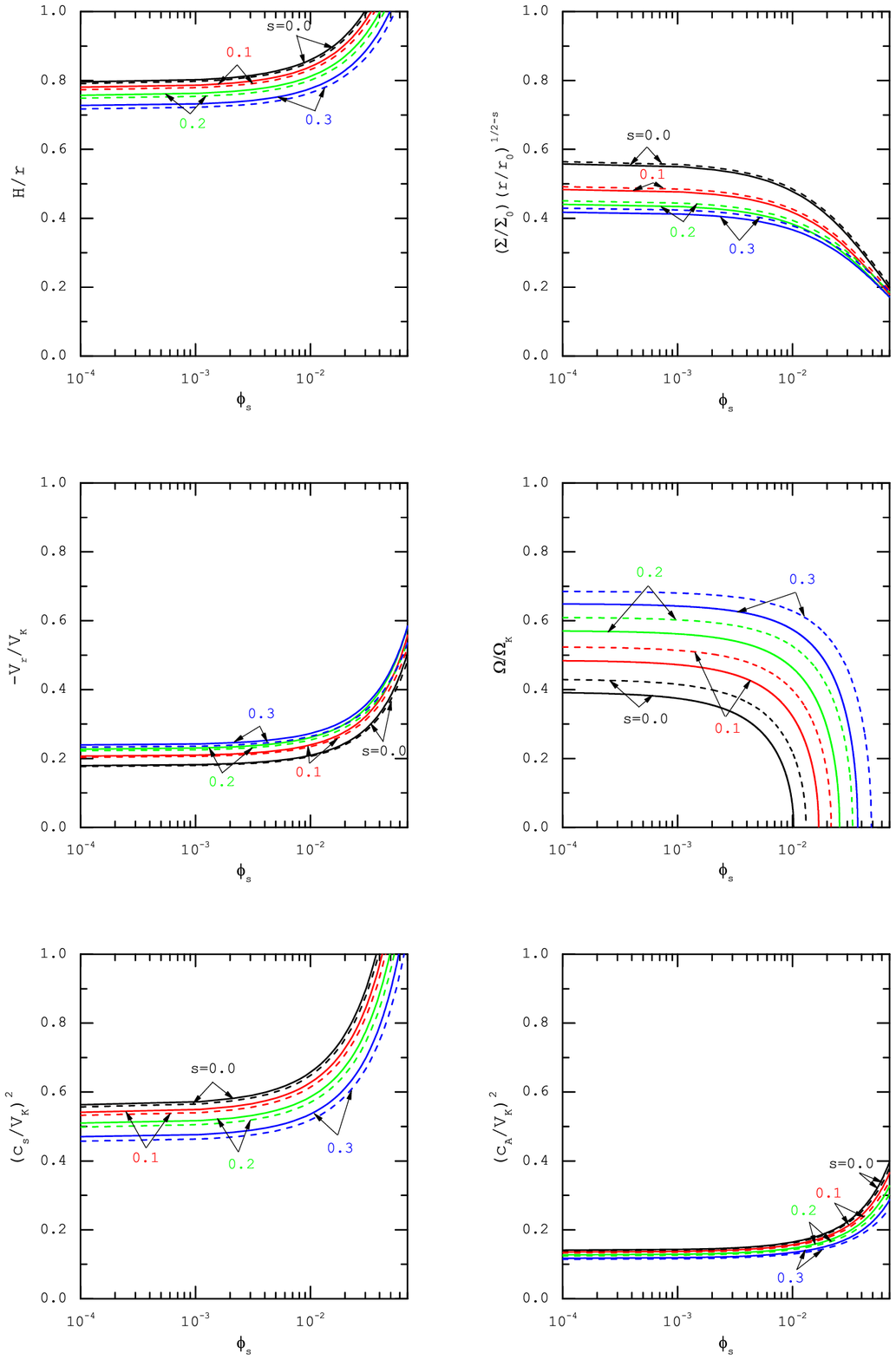}
\caption{Similar to Fig. \ref{fig:f2}, but for $\ell=0.0$. The solid and dashed curves are the cases with and without the radial viscosity, respectively.}\label{fig:f6}
\end{figure*}

As we mentioned before, the radial viscous force also depends on $\ell$ (see equation (\ref{eq:radial-force2})). In Fig. \ref{fig:f6}, the disc properties are investigated for $\ell=0.0$. The other input parameters are similar to Fig. \ref{fig:f2}. The solid curves correspond to the discs with the radial viscosity, while the dashed curves show the cases without this type of viscosity. The trend of disc quantities is the same as that shown in Fig. \ref{fig:f2}. One can see that the influence of wind and of radial viscosity become inconspicuous if the wind is assumed to be non-rotating.

\begin{figure*}
\includegraphics[scale=1.0]{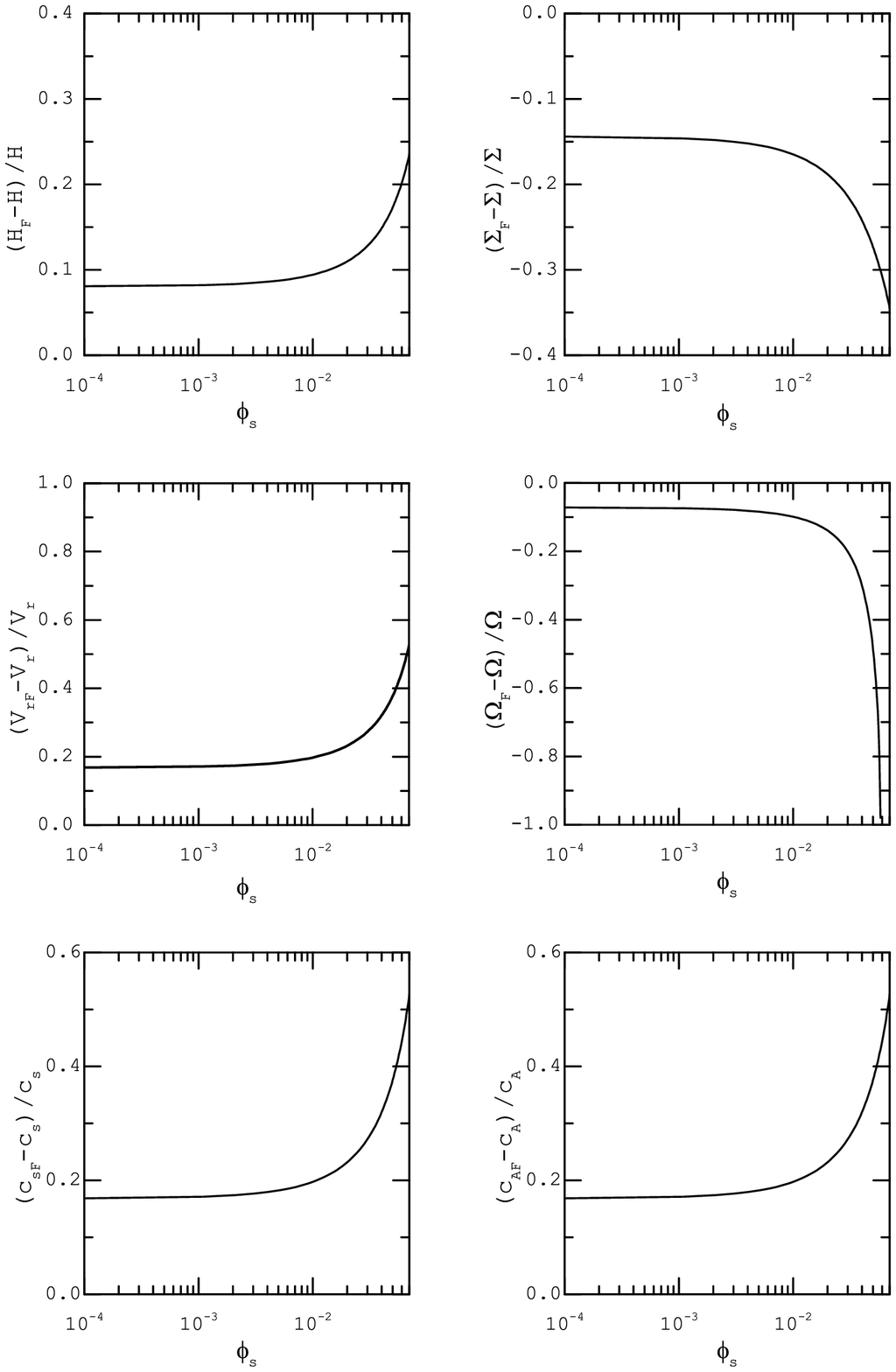}
\caption{The relative difference of disc parameters for $\zeta=1.0$, $\alpha=0.2$, $\beta=0.5$, $\ell=1.0$, and $s=0.3$. The variables in the presence of radial viscosity have a subscript F. The variables without subscript are related to a disc without the radial viscosity. }\label{fig:f7}
\end{figure*}

In Fig. \ref{fig:f7}, we illustrate the relative difference of disc parameters for $\beta=0.5$, $\zeta=1.0$, $s=0.3$, and $\ell=1.0$. The subscript 'F' reflects the fact that the quantity is obtained in the presence of radial viscosity. When the relative difference of a variable becomes negative, this means that the presence of radial viscosity causes this variable to reduce (see the profiles of surface density and rotational velocity). One can see that the relative difference of all variables is nearly constant in the small$-\phi_s$ limit. For high values of $\phi_s$, the relative difference becomes more, especially for the angular velocity. We find that the relative difference of disc thickness grows exponentially. Such a trend is seen for the relative difference of the speed sound, the Alfv\'{e}n velocity, and of course the infall velocity. However, the relative difference of the angular velocity and the surface density is exponentially decaying.

\begin{figure}
\includegraphics[scale=0.5]{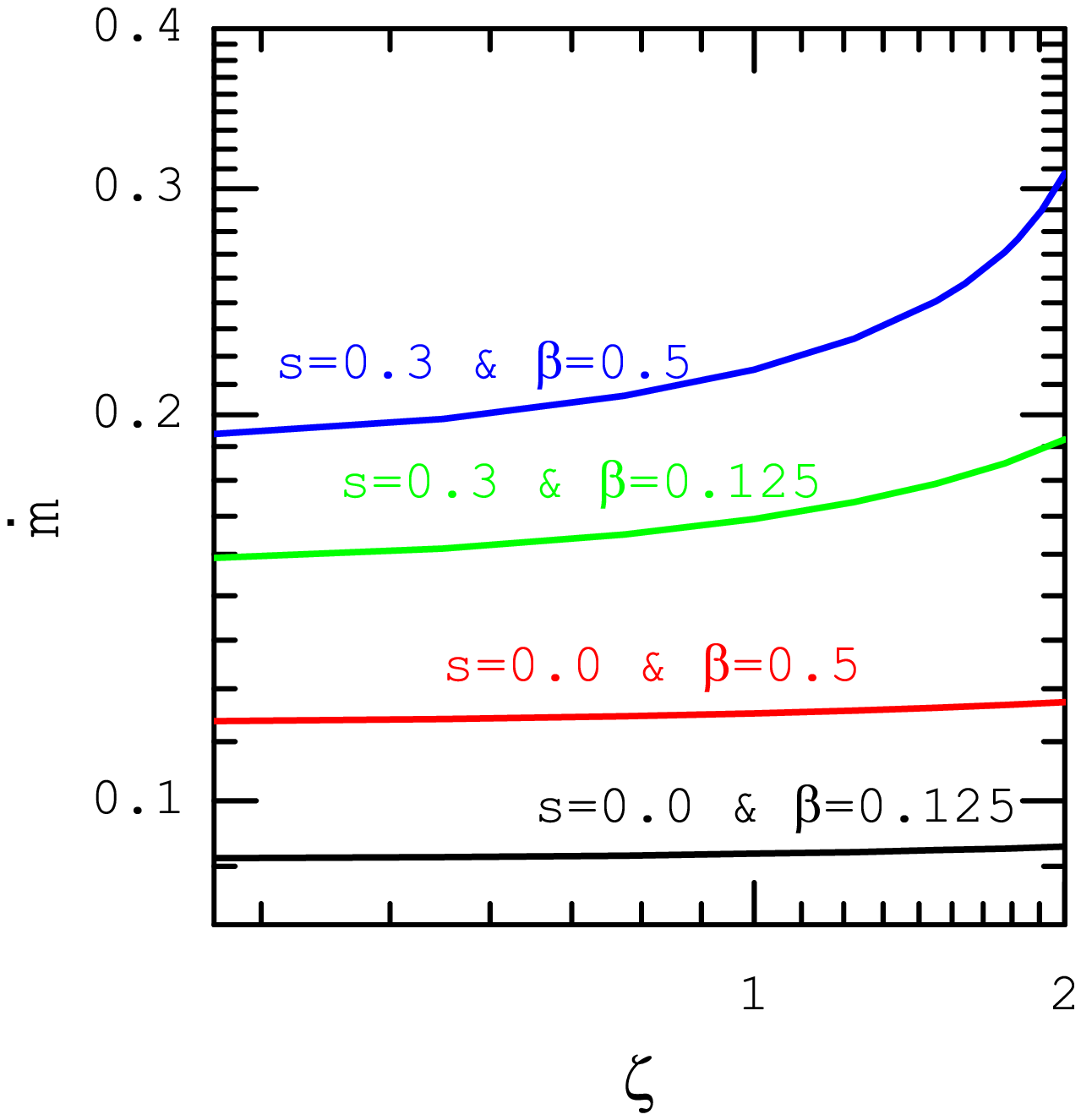}
\caption{Profile of the non-dimensional mass accretion rate $\dot{m}$ versus $\zeta$ for $r=r_0$, $\Sigma/\Sigma_0=1/2$, $\phi_s=0.001$. Each curve is labeled by the adopted exponent $s$ and the parameter $\beta$.}\label{fig:f8}
\end{figure}

According to the definition of accretion rate (see equation (\ref{eq:accretion-rate})), the accretion rate depends on the surface density and the radial velocity. As we found earlier, the radial viscosity leads to the reduction in the disc surface density. However, the disc materials out of this type of viscosity have a faster infall velocity. These findings are not adequate to illustrate the importance of radial viscosity in the accretion rate. However, we explore the role of $\zeta$ in the non-dimensional mass accretion rate $\dot{m}$ in Fig. \ref{fig:f8}. Here, we show profile of $\dot{m}$ as a function of $\zeta$ for
$r=r_0$, $\Sigma=(1/2) \Sigma_0$, and $\phi_s=0.001$. In the case of no wind, i.e., $s=0$, the accretion rate is almost unchanged. For the moderate-intensity winds, however, the accretion rate enhances as $\zeta$ increases. This means that the changes in infall velocity is more than the reduction of surface density. Note that the magnetic field strength can improve the enhancement of the accretion rate.


\section{Summary and Discussion}

The outflows and the radial viscosity can significantly affect the dynamics of advection-dominated discs. We have presented self-similar solutions for the advection-dominated discs  taking into account the radial viscosity and outflows in the presence of a toroidal component of the magnetic field and the thermal conduction. We also assumed that outflows can carry away some fractions of the disc material, the angular momentum, and the energy. Motivated by the results of numerical simulations and the observational evidences, we then prescribed the accretion rate as a power-law function of the radial distance. The power-law exponent indicates the strength of outflows and the observational results show that this index is around 0.3-0.4 \citep[e.g.,][]{Yuan2003}. We found that the outflow strength, the ratio of viscosities, the viscosity parameter, and the magnetic field strength are key parameters that strongly affect the effectiveness of the radial viscosity.

We can now summarize our main findings:
\begin{itemize}
\item The radial viscosity leads to a lower disc surface density. The reduction in the disc surface density is more significant as the wind gets stronger. In addition to the radial viscosity and the wind strength, the toroidal magnetic field and the viscosity parameter also contribute to the surface density reduction.
\item Our self-similar solutions show that the rotational velocity is always sub-Keplerian. The disc rotates with a slower rate as either the ratio of radial to azimuthal viscosities or the viscosity parameter increases. When a more mass is extracted by the winds, on the other hand, the rotational velocity increases. At a specific thermal conduction coefficient, the rotation of flow vanishes and the disc material has a purely radial motion. Under a purely radial motion, one can expect that the accretion rate at this specific thermal conduction coefficient (or even its higher value) increases.
\item In the presence of radial viscosity, the infall occurs with a higher velocity. We found that an increase in the wind strength, the magnetic fields, or the viscosity parameter can significantly affect on this result.
\item Considering the radial viscosity yields higher values of sound speed and of Alfv\'{e}n velocity. But the stronger winds cause these speeds to decrease.
\item Although the outflows reduce the disc thickness, the radial viscosity and magnetic field lead to a thicker disc.
\end{itemize}

In summary, we found that the surface density of an advection-dominated disc decrease because of the radial viscosity. The angular momentum removal in the presence of the radial viscosity is also more significant. This fact leads to a higher radial velocity. When the infall material moves with a faster velocity, the accretion rate may increase and therefore the disc density reduces. Hence, it is expected that the disc lifetime is shorter than the lifetime of a disc with only the azimuthal viscosity. This result could explain why the observational data show shorter lifetime for real discs. On the other hand, an additional viscosity yields lower rotational speed which implies a thicker disc. We also know that the radial viscosity acts as a heating agent. As expected, this additional heating mechanism can increase the disc temperature and thus the sound speed. But the stronger winds may reduce the disc temperature because such winds remove higher value of energy from disc. As suggested by \cite{Bu2009}, such a reduction in temperature due to outflows could explain difference between theoretical and observational temperatures.


\section*{Acknowledgements}
We would like to than referee for a constructive report that helped us to improve themanuscript. This work has been supported financially by Research Institute for Astronomy \& Astrophysics of Maragha (RIAAM) under research project No. 1/5237-62.

\bibliographystyle{mnras}
\bibliography{reference} 



\bsp	
\label{lastpage}

\end{document}